\documentclass[amssymb, showpacs, showkeys, aps, prl, superscriptaddress, twocolumn, longbibliography]{revtex4-2}
\usepackage[colorlinks,bookmarks=false,citecolor=blue,linkcolor=red,urlcolor=blue]{hyperref}

\usepackage[all]{hypcap}
\usepackage{xcolor}
\usepackage{slashed}
\usepackage{graphicx}
\usepackage{amsmath}
\usepackage{latexsym}
\usepackage{epstopdf}
\usepackage{amsmath}
\usepackage{enumitem}
\usepackage{amssymb,amsmath}
\usepackage{multirow}

\usepackage{mathtools}
\usepackage{bbm}
\usepackage{bm}
\usepackage{amsfonts}
\usepackage{amssymb}
\usepackage{calligra}
\usepackage{calrsfs}
\usepackage{makecell}
\usepackage[normalem]{ulem}
\usepackage[USenglish,american]{babel}
\usepackage{physics}
\usepackage{braket}

\newcommand{\be}{\begin{equation}}
\newcommand{\ee}{\end{equation}}
\newcommand{\ba}{\begin{eqnarray}}
\newcommand{\ea}{\end{eqnarray}}
\newcommand{\bs}{\begin{split}}
\newcommand{\es}{\end{split}}

\definecolor{darkblue}{rgb}{0.0,0.0,0.4}
\definecolor{darkgreen}{rgb}{0.0,0.4,0.0}
\definecolor{darkred}{rgb}{0.6,0.0,0.0}
\newcommand{\cred}[1]{{\color{red}{#1}}}

\usepackage{comment}

\begin{document}

\title{\cred{TL: Beyond scalar tunability in molecular quantum gases}}
\title{\cred{MC: Droplet phases of fully anisotropic microwave-dressed polar molecules in a trap}}
\title{Equilibrium and non-equilibrium phases of microwave-dressed polar molecules in the absence of rotational symmetries}
\title{Equilibrium and non-equilibrium phases of microwave-dressed\\polar molecules beyond rotational symmetries}

\author{Matteo Ciardi}
\email{matteo.ciardi@tuwien.ac.at}
\affiliation{Institute for Theoretical Physics, TU Wien, Wiedner Hauptstraße 8-10/136, 1040 Vienna, Austria}

\author{Andreas Schindewolf}
\email{andreas.schindewolf@tuwien.ac.at}
\affiliation{Vienna Center for Quantum Science and Technology,
Atominstitut, TU Wien, Stadionallee 2, 1020 Vienna, Austria}

\author{Tim Langen}
\email{tim.langen@tuwien.ac.at}
\affiliation{Vienna Center for Quantum Science and Technology,
Atominstitut, TU Wien, Stadionallee 2, 1020 Vienna, Austria}

\author{Thomas Pohl}
\email{thomas.pohl@itp.tuwien.ac.at}
\affiliation{Institute for Theoretical Physics, TU Wien, Wiedner Hauptstraße 8-10/136, 1040 Vienna, Austria}

\begin{abstract}
{
Recent experiments on molecular droplets have opened a new frontier of self-organization in strongly dipolar quantum matter. Microwave-dressing of polar molecules permits to tune both the strength and the angular structure of long-range interactions, potentially promoting a rich spectrum of quantum phases, from superfluid droplets with varying geometry and insulating or supersolid droplet arrays to strongly correlated crystals of individual molecules. Using path-integral Monte Carlo simulations of large molecular ensembles, we demonstrate that experimentally observed droplet arrays emerge as a metastable non-equilibrium state from the quenching of a gas-droplet phase transition under entirely broken rotational symmetry of the microwave-induced interaction potential. We moreover find that a crystalline phase of molecules, predicted for antidipolar interactions, is absent under conditions of recent experiments. This is traced back to the lack of angular symmetry in currently employed microwave-dressing, which qualitatively reshapes the many-body energy landscape and cannot be captured by effective scalar interaction parameters.
Our results provide the first direct comparison of ab initio simulations and experiments and establish interaction anisotropy as a key aspect of molecular quantum gases.
}
\end{abstract}

\maketitle
\begin{figure}[t!]
  \begin{center}
    \includegraphics[width=0.95\linewidth]{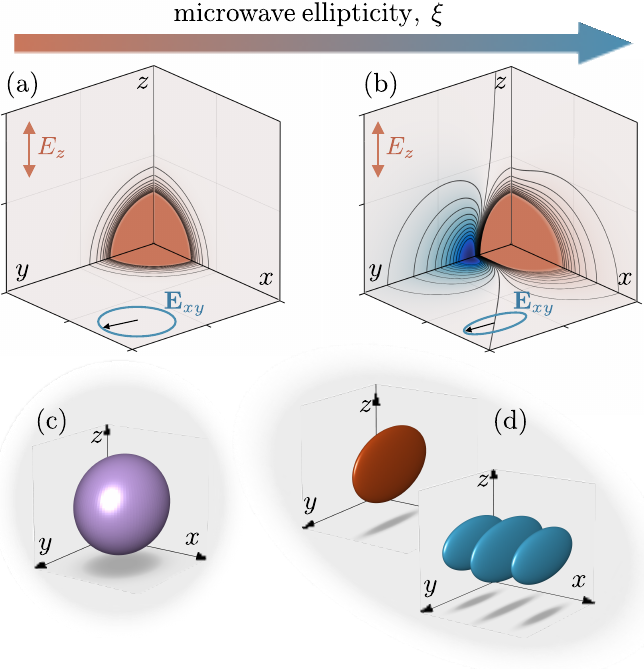}  
    \caption{(a,b) Contour plots of the interaction potential between two polar molecules, dressed by a linearly ($E_z$) and elliptically polarized (${\bf E}_{xy}$) microwave field. (a) The potential is isotropically short-ranged and repulsive under circular driving ($\xi=0$). (b) It acquires a long-range attractive well along the $y$-axis and dipolar repulsion along the $x$-axis for finite ellipticities $\xi>0$. (c) For $\xi=0$, the isotropic short-range repulsion results in a simple Thomas-Fermi profile of trapped BECs. (d) For $\xi>0$, the induced long-range repulsion and attraction in the $(x-y)$-plane stabilizes single quantum droplets and generates meta-stable droplet arrays, under conditions of recent experiments \cite{Zhang2026}. 
    }
    \label{fig1}
  \end{center}
\end{figure}

The formation of self-organized structures and crystalline order is among the most striking consequences of long-range interactions in quantum matter. Ultracold polar molecules have long promised access to such phenomena, but only recent advances in microwave shielding \cite{Karman2018, Lassabliere2018, Anderegg2021, Schindewolf2022, Deng2025, Karman2025, Yuan2025, Schindewolf2026} have enabled experiments in the strongly degenerate regime, and in particular molecular Bose-Einstein condensates (BECs) \cite{Bigagli2024, Shi2025}. This progress has triggered a surge of theoretical and experimental activity, suggesting a rich spectrum of self-organization phenomena, including molecular quantum droplets, superfluid membranes, density-wave supersolids, and molecular crystals \cite{Zhang2026, Langen2025, Jin2025, Ciardi2025, Zhang2025d, Deng2025, Zhang2025, Zhang2025d, Shi2026, Wang2026, Polterauer2025, Baillie2026, ArnoneCardinale2026, Wang2026, Zampronio2026}.
In atomic dipolar Bose gases, the formation of different phases can be understood in terms of a small number of scalar interaction parameters, such as the s-wave scattering length and dipolar length \cite{Bottcher2021, Chomaz2022} of the atomic interaction. Cold polar molecules offer an additional degree of control, as microwave dressing gives rise to distinct long-range interactions with a more complex and broadly tuneable angular structure \cite{Schindewolf2026} (see, e.g., Fig.~\ref{fig1}). While the angular features of the induced interaction potential are crucial for the stability of molecular BECs, their effects on the resulting many-body physics have remained largely unexplored.

Here, we address this open question using path-integral Monte Carlo (PIMC) simulations of more than $10^3$ molecules, matching the conditions of recent experiments~\cite{Zhang2026} that reported self-bound quantum droplets and  regular droplet arrays of microwave-dressed NaCs molecules. Detailed comparisons between experiment and theory demonstrate that the observed droplet arrays emerge naturally during the preparation protocol of Ref.~\cite{Zhang2026}. However, we find that such ordered structures represent non-equilibrium states that result from the quench dynamics at a ground-state phase transition between the extended gas phase and an anisotropic compact quantum droplet [Fig.~\ref{fig1}(b)--(d)]. In the strong-interaction regime, where the crystallization of individual molecules has been anticipated \cite{Zhang2026} and specifically predicted for rotational symmetric microwave dressing \cite{Ciardi2025}, we do not find any evidence of a strongly correlated phase. All these findings are directly traced back to the complete rotational symmetry breaking of the dipole-dipole interaction \cite{Bigagli2024,Zhang2026,Shi2025}. Our results thus highlight the crucial role of the tuneable angular structure of molecular long-range interactions that substantially enriches the many-body physics of dipolar quantum matter and cannot be captured by a simple scalar parametrization of the interaction strength. 

We study an ensemble of $N$ bosonic molecules with mass $m$, as described by the Hamiltonian 
\begin{equation} \label{eq:hamiltonian}
\hat{H} = \sum_{i=1}^{N} \left[ \frac{\hbar^2}{2 m} \nabla_{{\bf r}_i}^2 + V({\bf r}_i) \right] + \sum_{i<j} U({\bf r}_i-{\bf r}_j).
\end{equation}
The particles are confined in a harmonic potential $V({\bf r})$ with individual trapping frequencies $\omega_x$, $\omega_y$, and $\omega_z$ along each of the cartesian directions. The interaction potential $V({\bf r})$ between two molecules can be engineered and tuned via external fields. In a number of recent experiments \cite{Anderegg2021, Schindewolf2022, Chen2023, Lin2023, Bigagli2023, Bigagli2024, Yuan2025, Zhang2026, Biswas2026}, microwave fields \cite{Buchler2007, Micheli2007, Gorshkov2008, Cooper2009, Huang2012, Karman2018, Lassabliere2018} have been used to engineer a repulsive van der Waals potential $V({\bf r})\sim1/r^6$ in order to shield the molecules from short-range collisions and associated losses at short distances \cite{Bause2023}. At the same time, microwave dressing generates long-range dipole-dipole interactions, whose strengths and shape can be controlled via the applied fields \cite{Deng2023, Karman2025, Deng2025, Jozwiak2026}. Applying simultaneously a linearly and an elliptically polarized microwave field induces a combination of dipole-dipole interactions, whose strengths scale as $d_z^2\propto 1-3z^2/r^2$ and $d_{xy}^2\propto (x^2-y^2)/r^2$, respectively, with the distance vector ${\bf r}=(x,y,z)$ between the two molecules. 

A recent experiment \cite{Zhang2026} used a fine-tuned combination of linearly and elliptically polarized fields to cancel the strength $d_z$ of the common dipole-dipole interaction. This generates a long-range potential that can be expressed as \cite{Karman2025,Deng2025,Zhang2026,Biswas2026} 
\begin{equation} \label{eq:interaction}
    U(\textbf{r}) = \frac{C_6}{r^6} + \frac{C_3 }{r^3} \sin (2\xi)\frac{x^2-y^2}{r^2}, 
\end{equation}
and lacks rotational symmetry around any of the three cartesian axes. By tuning the ellipticity $\xi$ of the elliptically polarized field, one can control the strength and the opposing sign of the long-range dipole-dipole along the $x$- and $y$-axis, respectively, while maintaining the short-range repulsion in the $z$-direction. As we will see below, this strong anisotropy has crucial consequences for the resulting quantum phases of the system.

We study the system via exact numerical simulations, using path integral Monte Carlo (PIMC) simulations \cite{ceperley1995}, which have been applied successfully to dipolar bosons \cite{Micheli2007,Jain2011, Macia2011, Saito2016,  Bottcher2019b, Boninsegni2021, Langen2025,Ciardi2025,Zhang2025, Cardinale2025}. Observables are hereby sampled at finite temperature  from a Monte-Carlo integral over a set of discrete imaginary-time trajectories, i.e.\ world lines  associated to each particle.

We simulate a large number of particles ($N=1200$) which locates us in the experimental regime. All simulations are performed for the typical temperatures and  interaction coefficients ($C_3$, $C_6$) of the experiment \cite{Zhang2026} (see End Matter).  We also consider the asymmetric trapping potential used in \cite{Zhang2026}, with the strongest confinement along the $x$-axis with  $\omega_x/2\pi =15$\,Hz, while $\omega_y/2\pi =27$\,Hz and $\omega_z/2\pi =53$\,Hz. Keeping these parameters fixed, we study the properties of the molecular Bose-Einstein condensate as we vary the strength of the dipole-dipole interaction in Eq.~(\ref{eq:interaction}) by changing the microwave ellipticity $\xi$.

For $\xi = 0$, the dipolar term in Eq.~\eqref{eq:interaction} vanishes, and the remaining van der Waals interaction is isotropically repulsive. The two-body scattering is, in this case, described by an effective zero-range potential with a scattering length of $\sim 2100\,a_0$. Indeed, our PIMC simulations for $\xi=0$ agree accurately with the simple prediction of the Thomas-Fermi approximation using this scattering length (see End Matter).

Increasing the microwave ellipticity, $\xi$, induces dipolar interactions, whose effects were probed in \cite{Zhang2026} through time-of-flight measurements.
Small ellipticities $|\xi|\lesssim3^\circ$, were observed to preserve the typical expansion dynamics of weakly dipolar BECs, as reported previously for magnetic atoms \cite{Stuhler2005}. While our PIMC simulations do not permit to explore real-time dynamics, we find that the size and geometry of the equilibrated condensate remain virtually unaffected by the dipolar interaction for $|\xi|\lesssim3^\circ$. This is consistent with the weak effect of dipole-dipole interactions on the ensuing expansion dynamics observed in \cite{Zhang2026}.  

Beyond $\xi\gtrsim3^\circ$, however, our simulations reveal a structural transition of the trapped molecular cloud, which coincides with the formation of  self-bound droplet states seen in the time-of-flight measurements \cite{Zhang2026}. Perhaps most strikingly, at positive values $\xi\gtrsim3^\circ$, the experiment found a fragmentation of the molecular condensate into small droplet arrays, in which the resulting structure was observed to depend on the ramp speed of increasing the ellipticity from $\xi=0$ to its finite value $\xi\gtrsim3^\circ$. Notably, in this regime, our simulation results also feature an increased sensitivity to the initial state that is used to seed the PIMC sampling. 

\begin{figure*}[t!]
  \begin{center}
  \includegraphics[width=\linewidth]{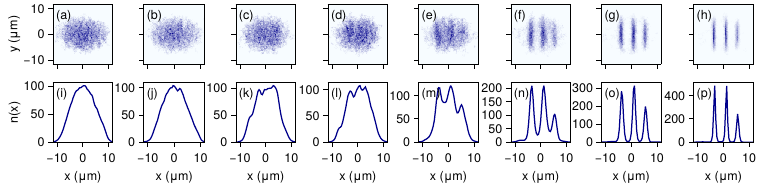}
    \caption{Molecular density profiles at $T=2\:$nK obtained by consecutively increasing the microwave ellipticity, $\xi$, from $2.25^\circ$ (a,i) to $4.0^\circ$ (h,p) in steps of $0.25^\circ$. The top row, (a) - (h) depicts the projected density in the $x - y$ plane, while the bottom row shows the density profile integrated along the $y$-axis. Starting from a Thomas Fermi profile (BEC phase), the condensate continuously transitions into an array of disc-shaped quantum droplets upon gradually increasing $\xi$.}
    \label{fig2}
  \end{center}
\end{figure*} 

In order to mimic the slow-ramp preparation protocol of the experiment in the PIMC simulations, we adopt the following procedure. As in the experiment, we first equilibrate the system at $\xi=0$, to obtain our initial configuration, corresponding to the simple Thomas-Fermi profile that reflects the geometry of the trap. Next, we increase the ellipticity in small steps of $\xi=0.25^\circ$ and equilibrate the systems at these consecutive values of $\xi$ using the final state of the previous step as the initial configuration for the PIMC sampling. While performing such consecutive PIMC simulations for our large particle number $N=1200$ is numerically demanding, it permits to avoid sharp quenches of the ellipticity from $\xi=0$ to large values of $\xi$. We find the latter to cause the disintegration of the initial cloud into small molecular clusters, consistently with the experimentally observed behavior for faster ramps of the ellipticity. 

The Monte Carlo results of our slow-ramp protocol are shown in Fig.~\ref{fig2}, where we show the density profile in the $x$--$y$ plane and the integrated density $n_x(x)$ along the $x$-axis. While the molecular BEC initially remains affected only marginally by the increasing dipole-dipole interaction, it develops pronounced density modulations as the ellipticity crosses a critical value $\xi\approx3^\circ$. Upon further increasing $\xi$, the modulations become more pronounced and eventually form an array of separated droplets along the $x$-axis. As shown in Fig.~\ref{fig2}(m--p), the droplets become increasingly sharp at large values of $\xi$, giving rise to disc-shaped structures that lie in the $y$--$z$ plane and are stacked along the $x$-direction. 

\begin{figure}[b!]
  \begin{center}
  \includegraphics[width=\linewidth]{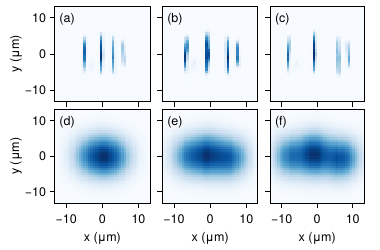}
    \caption{Time-of-flight dynamics of metastable droplet arrays. (a) Simulated density in the $x-y$ plane for $\xi = 4.7^\circ$ in the trap at $t=0$.  Panels (b) and (c) show the calculated density after time-of-flight expansion at later times $t=10\:$ms (b) and $t=25\:$ms (c), respectively. Panels (d)-(f) show the corresponding density profiles in the $x-y$ plane, incorporating the finite imaging resolution of $3\:\mu$m in the experiment \cite{Zhang2026}.}
    \label{fig3}
  \end{center}
\end{figure}

While the predicted structural transition at $\xi\sim3^\circ$ is consistent with the behavior found experimentally in \cite{Zhang2026}, the geometry of the observed droplet arrays is markedly different from our theoretical results. Secondly, droplet arrays were not observed directly as an equilibrium phase of the confined dipolar BEC, but only after its release from the trap with more pronounced density modulations developing in the course of the time-of-flight expansion. The first difference is readily traced back to the finite resolution $\Delta r\approx 3\,\mu$m of the imaging setup in \cite{Zhang2026}. Convoluting the theoretical density distribution with a corresponding Gaussian point spread function indeed causes a significant broadening of the otherwise sharp density peaks predicted by simulations. This is demonstrated in Fig.~\ref{fig3}(a) and (d) where we compare the calculated density to the predicted image after convolution. Since $\Delta r$ is comparable to the separation between the disc-shaped droplet, the sharp density peaks are blurred entirely and no longer distinguishable in the calculated image.

In order to study the subsequent time-of-flight expansion, we adopt a simplified semi-classical approach. For the considered large values of $\xi$, the individual droplets are spatially well separated, without any exchange of particles between them. We may, hence, approximate their dynamics as a form-stable collective motion of each droplet due to their mutual interaction. The details of the calculation are in the End Matter, and the results are exemplified in Fig.~\ref{fig3}(a--c), showing the molecular density and the correspondingly expected images for  different expansion times. We find that the collective long-range repulsion between droplets along the $x$-axis is sufficiently strong to expand the droplet array on ms time scales. Indeed, this repulsion eventually makes the droplet structure visible in the blurred time-of-flight images in accordance with the experimental observations, as seen in Fig.~\ref{fig3}(e--f).

\begin{figure}[t!]
  \begin{center}
  \includegraphics[width=\linewidth]{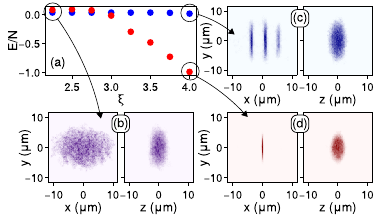}
    \caption{Stability of Thomas-Fermi solution, single droplet, and droplet array. Panel (a) compares the energy, $E$, for the states of Fig.\ref{fig2} (blue points) with PIMC simulations, initialized from a single compact disc at an ellipticity $\xi=4^\circ$ and subsequent stepwise decrease of $\xi$ (red point). For small ellipticities, $\xi\lesssim3^\circ$, both preparation protocols converge to the Thomas Fermi profile (b), expected for weakly dipolar BECs. For large ellipticities, $\xi\gtrsim3^\circ$, the experimentally observed droplet array (c) emerges as a metastable non-equilibrium phase with a higher total energy, while a single droplet (d) represents the ground state of the system. Colors in the density plots are chosen to match Fig.~\ref{fig1}. 
    }
    \label{fig4}
  \end{center}
\end{figure} 

Having established the close correspondence between the droplet arrays observed experimentally and numerically, we can use the simulations to explore the underlying physical mechanism. Previous work on weakly interacting atomic dipoles has found similar behavior in the form of density-wave supersolid phases, emerging due to a roton-like instability or ground-state phase transitions upon increasing the relative strength of dipole-dipole interactions \cite{Boettcher2019, Tanzi2019, Chomaz2019, Bottcher2021, Chomaz2022}. In order to clarify whether a similar mechanism may be at work in the present case, we have performed additional simulations using a different preparation protocol.

The preparation procedure is to initialize the simulations from a single disc-shaped cloud that is closer to the geometry of the self-confined droplet solution and equilibrate the system for different values of $\xi$. The results are summarized in Fig.~\ref{fig4}, where we compare the total energy of the molecules as a function of $\xi$ obtained from the two different preparation protocols. When initializing the system in the disc-shaped droplet state, we find that the molecules remain compact (see Fig.~\ref{fig4}(d)) and do not fragment into the droplet array found from the adiabatic preparation starting from the Thomas-Fermi solution at $\xi=0$ (cf.\ Fig.~\ref{fig2} and Fig.~\ref{fig4}(b)). Remarkably, we find that the energy of the single-droplet solution features a lower energy, indicating that the droplet array observed in our simulations as well as in the experiment emerges as a metastable phase due to the preparation protocol starting from a much lower density.

These results indicate that the observed droplet array is neither the result of an equilibrium phase transition to a density-wave supersolid, 
nor does it emerge from a roton instability of the BEC, in contrast to the behaviour of dilute dipolar atomic gases. Instead, the physical mechanism behind the formation of the density-wave state and the single-disc geometry are both related to the transition to a self-confined droplet phase in free space, appearing at $\xi\approx3^\circ$. For smaller values of $\xi$, the molecular cloud fully delocalizes in free space and fills the trap following the Thomas-Fermi profile with only minor corrections from the long-range dipole interactions, as discussed above. The geometry of the self-bound droplet phase for $\xi\gtrsim3^\circ$, on the other hand, is much more compact along the $x$-direction than the Thomas-Fermi solution for $\xi\lesssim3^\circ$. Crossing the critical value of $\xi$ triggers a quench, which inevitably leads to fragmentation of the extended cloud into smaller droplets stacked along the $x$-direction [see Fig.~\ref{fig4}(c)]. The dipolar repulsion between the fragments then results in a regular droplet array and renders this state metastable, even as $\xi$ is increased further. 

On the contrary, when preparing the molecular BEC in a compact single-disc state at large values $\xi>3^\circ$, the system is enabled to settle into its ground state, which resembles the free-space droplet phase with some corrections from the external confinement. 
The pronounced anisotropy of the resulting droplet reflects the angular structure of the interaction potential. Long-range repulsion along the $x$-axis favors strong compression of the cloud in this direction, minimizing potential energy. Long-range attraction along $y$ and merely short-range repulsion along $z$ correspondingly favor an extension of the droplet along these two axes. The ground state, therefore, assumes a disk-like geometry that is narrow along $x$ and extended within the $y–z$ plane. 
Upon decreasing $\xi$ below the critical value, the system expands into the trap and settles back into the low-density Thomas-Fermi solution, dominated by the short-range interaction.  

The disc-shaped geometry of the solutions in the dipolar regime, $\xi>3^\circ$, is reminiscent of that predicted in molecular BECs dressed with a single circularly polarized microwave, which produces an effective anti-dipolar interaction. In that case, the resulting interaction potential features in-plane rotational symmetry and stabilizes disc-shaped quantum droplets in a trap or in free space \cite{Langen2025,Jin2025,Ciardi2025}. 
This interaction leads to exotic phenomena, including the formation of a two-dimensional crystalline layer of individual molecules in the strongly dipolar regime, even when a two-body bound state is absent. However, since in Ref.~\cite{Zhang2026} dual-microwave dressing was employed to realize long BEC lifetimes, which featured the potential of Eq.~(\ref{eq:interaction}) with broken rotational symmetry, it is important to clarify whether this can promote similar crystalline phases. 
Surprisingly, we find no evidence of this behaviour, even at such large interaction strengths that would place molecules with single-microwave dressing deep into the crystalline regime (see End Matter). 

\begin{figure}[t!]
  \begin{center}
  \includegraphics[width=\linewidth]{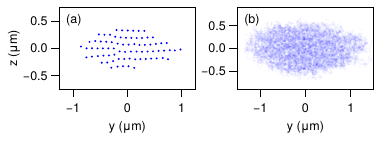}
  \caption{Classical vs. quantum ground state in the strongly dipolar regime. Panel (a) shows the result of a classical zero-temperature Monte Carlo simulation of $N=75$ molecules for $\xi=8^\circ$. The classical treatment predicts an ordered state of self-bound molecular strings, with weak interactions between the multimers. However, the equivalent result of a PIMC sampling (b) does not preserve this classical order and instead yields a structureless superfluid droplet.}
    \label{fig5}
  \end{center}
\end{figure}

In order to understand this observation, we have performed classical Monte Carlo simulations in the strong-interaction regime, for $\xi=8^\circ$. As shown in Fig.~\ref{fig5} for a smaller system of $N=75$ molecules, the classical particles form a single-layer in the $y$--$z$ plane and align along the $y$-axis on tightly bound chains with a lattice constant corresponding to the deep potential minimum shown in Fig.~\ref{fig1}(b). While the classical simulations also yield ordering between adjacent chains in the transverse direction, the interaction between the chains is comparably weak and dominated by the pure short-range repulsion along the $z$-direction [see Fig.~\ref{fig1}(b)]. In contrast with the robust two-dimensional crystal observed for antidipolar interactions, hence, we obtain a collection of weakly repulsive molecular strings. The latter is far more sensitive to perturbations, and in particular thermal and quantum fluctuations. This is demonstrated in Fig.~\ref{fig5}(b), where we show the results of a corresponding PIMC simulation using the classical state of Fig.~\ref{fig5}(a) as the initial configuration. Both the two-dimensional in-plane alignment and the crystalline order disappear in the quantum case. 
The anisotropic interaction is insufficient to maintain the classical order and the effectively one dimensional strings, that result from the strong anisotropy, do not survive in the presence of quantum fluctuations \cite{Cazalilla2011}, and instead delocalize to form an extended superfluid droplet. It is notable that molecular strings in free space are conjectured to be stable once the potential acquires a two-body bound state \cite{Wang2026}.

In summary, we have performed quantum Monte Carlo simulations of more than $10^3$ microwave-dressed molecules under conditions of recent experiments. The simulations reproduce the experimentally observed phases, from asymmetric droplets to regular arrays of quantum droplets. However, our results show that the latter emerges as a non-equilibrium metastable state from a fragmentation process as the system is quenched through the gas-droplet phase transition by varying the applied microwave field. The peculiar structure of stacked pancake-shaped condensates results from the entirely broken rotational symmetry of the molecular interaction featuring short- and long-range repulsion as well as long-range attraction along each of the three cartesian axes. The strong asymmetry is, moreover, found to inhibit crystalline phases, which are otherwise predicted for anti-dipolar molecules \cite{Ciardi2025} that feature rotationally symmetric long-range attraction. 

Our findings apply directly to recent experiments that realize the interaction potential of Eq.~(\ref{eq:interaction}) via accurately fine-tuned microwave fields. However, other quantum phases, such as ground-state droplet arrays or two-dimensional droplet crystals, strongly correlated crystals and chains of molecules as well as quantum phase transitions between insulating structures and supersolid phases, may emerge for different trap geometries and angular dependencies of molecular interactions \cite{Jin2025,Ciardi2025,Langen2025,Zhang2025,Zhang2025d,Shi2026,Wang2026,ArnoneCardinale2026}. Given the profound effects of such angular structure, demonstrated in this work, and their broad tunability via microwave dressing, ultracold molecules promise a unique platform that could broaden the scope of dipolar quantum matter and expand their already rich spectrum of quantum phases found for common static dipole-dipole interactions. We hope that the findings of this work initiate future theoretical and experimental investigations towards a systematic understanding of the role of angular symmetry of long-range interactions and its interplay with the trap geometry for the equilibrium and non-equilibrium behaviour of molecular quantum gases.

\begin{acknowledgments}
We thank Ian Stevenson and Tao Shi for valuable discussions. This work was supported by funding from the
Austrian Science Fund (FWF) through the SFB Qnnect (Grant No.\ 10.55776/F101200) and the cluster of excellence quantA (Grant No.\ 10.55776/COE1) and
the European Union (NextGenerationEU), by the SNSF
through the Swiss Quantum Initiative, and from the
European Research Council through the ERC Synergy
Grant "SuperWave" (Grant No.\ 101071882) and the ERC Starting Grants "NEWMAT" (Grant No.\ 949431) and "UltraMeDiQs" (Grant No.\ 101219560). 
\end{acknowledgments}

\bibliography{biblio}

\clearpage

\section{End Matter}

\subsection{Parameters and benchmarks} 

The parameters $C_3$ and $C_6$ can be fully determined based on experimental parameters. Following \cite{Zhang2026}, we have $C_3 = \frac{\hbar^2}{m} \sqrt{3} \times 53{,}100 \, a_0$, and $C_6 = \frac{\hbar^2}{m} (3200 \, a_0)^4$.

For the external potential, as mentioned in the text, we have the harmonic trapping
\begin{equation} \label{eq:external}
    V(\textbf{r}) = \sum_{d=1}^3 \frac{m \omega_d^2}{2} r_d^2 \,, 
\end{equation}
with $\omega_d = 2 \pi \times \{15, 27, 53\}\,\text{Hz}$ respectively. Taking the molecule mass $m_{\text{NaCs}} \approx 156$\,amu, these correspond to oscillator lengths of $2.072\,\mu$m, $1.545\,\mu$m, and $1.103\,\mu$m or $39{,}163 \, a_0$, $29{,}191 \, a_0$, and $20{,}835 \, a_0$, respectively.

\begin{figure}[h!]
  \begin{center}
  \includegraphics[width=\linewidth]{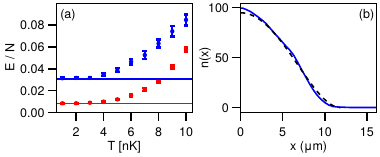}
    \caption{Thomas-Fermi energies and density compared to simulation results at $\xi=0$. (a): Blue points are simulations with the interaction of Eq.~\eqref{eq:interaction} at $\xi=0$, while red points are simulations of non-interacting bosons, all in the harmonic trap described by Eq.~\eqref{eq:external}. The blue and red solid lines represent the ground-state energy from Thomas-Fermi and non-interacting bosons, respectively. (b): Comparison between the density histogram measured in the simulation, integrated along $y$ and $z$, with the Thomas-Fermi prediction.}
    \label{fig:thomas-fermi}
  \end{center}
\end{figure}

Throughout this work, we discuss PIMC simulations with a large number of particles ($N=1200$). At $\xi=0$, the interaction is a simple potential of the form $C/r^6$, of known scattering length $a$, which can be compared directly with results from the Gross-Pitaevskii equation in the Thomas-Fermi limit (see, e.g.\ Ref.~\cite{Pitaevskii2016}). This result works as a benchmark for the PIMC code, and at the same time shows that the system still satisfies the dilute limit required by GPE, with a small correction to the density profile. This is shown in Fig.~\ref{fig:thomas-fermi}(b). We also show, in Fig.~\ref{fig:thomas-fermi}(a), the temperature dependence of the energy per particle in the range relevant to the experiment \cite{Zhang2026}.

\subsection{Potential rescaling}

In order to compare different potential shapes, it is useful to rescale the units of the potential to introduce a single dimensionless strength parameter, as done in \cite{Ciardi2025} and \cite{Schindewolf2026}. To this aim, at fixed $\xi$, we introduce the characteristic length $l = (C_6 / C_3 \sin (2\xi))^{1/3} $, which allows us to rewrite the potential \eqref{eq:interaction} as
\begin{equation}
    U(\textbf{r}) = C \left[ \frac{1}{r^6} + \frac{1}{r^3} \frac{x^2-y^2}{r^2} \right], 
\end{equation}
with the strength parameter taking the form
\begin{equation}
    C = \frac{m}{\hbar^2} \frac{(C_3 \sin (2\xi))^{4/3}}{C_6^{1/3}} \,.
\end{equation}

At $\xi=8^\circ$, with $C_3$ and $C_6$ introduced above, this leads to $C \approx 16$, which, for the antidipolar potential, is already fairly deep in the crystal phase even for $N=75$ particles.

\subsection{Semi-classical dynamics}

In a semi-classical approach which treats each droplet as a rigid, independent object, the forces between the droplets can be determined as follows. A single configuration in the PIMC simulation is a collection of $N$ discrete world lines. A world line represents a quantum particle and consists of $M$ positions for each of the $M$ discrete time slices used in the simulation. The forces between the clusters can then be determined from the world line coordinates ${\bf r}_i^{(m)}$ for each of the $i=1,...,N$ molecules, each represented by $m=1,...,M$ world line positions. Since the droplets are well separated, each world line can be uniquely assigned to a given droplet such that we can group the particles accordingly and calculate the total force 
\begin{equation}\label{eq:droplet_force}
    {\bf F}_{\alpha\beta} = \frac{1}{M} \sum_m \sum_{i \in \alpha, j \in \beta} {\bf f}_{ij}^{(m)} \,,
\end{equation}
between two droplets labeled by $\alpha$ and $\beta$ that contain the particles $i\in\alpha$ and $j\in\beta$, respectively. The microscopic force ${\bf f}_{ij}^{(m)}$ between two particles at ${\bf r}_i^{(m)}$ and ${\bf r}_j^{(m)}$ on the $m^{\rm th}$ time sheet is obtained from the interaction potential Eq.~(\ref{eq:interaction}). Treating the droplets as rigid bodies, we use ${\bf F}_{\alpha\beta}$ to propagate their center of mass ${\bf R}_\alpha=M^{-1} \sum_m \sum_{i \in \alpha} {\bf r}_{i}^{(m)}$ via classical equations of motion.

\end{document}